\documentstyle{l-aa}

\begin{document}
\newcommand{\be}{\begin{equation}}
\newcommand{\ee}{\end{equation}}
\newcommand{\bc}{\begin{center}}
\newcommand{\ec}{\end{center}}
\newcommand{\Hb}{H$\beta$}
\newcommand{\Ha}{H$\alpha$}
\newcommand{\DSS}{Digitized Sky Survey \footnote{Based on photographic
    data of the National Geographic Society -- Pal
    Observatory Sky Survey (NGS-POSS) obtained using the Oschin Telescope
    Palomar Mountain.  The NGS-POSS was funded by a grant from the Natio
    Geographic Society to the California Institute of Technology.  The
    plates were processed into the present compressed digital form with
    their permission.  The Digitized Sky Survey was produced at the Spac
    Telescope Science Institute under US Government grant NAG W-2166.}}
\newcommand{\IRAF}{IRAF\footnote{IRAF is distributed by the National Optical 
 Astronomy Observatories, which is operated by the Association of Universities 
 for Research in Astronomy, Inc., under cooperative agreement  with the 
 National Science Foundation.}}

 \thesaurus{03         
            (11.01.2; 11.19.1; 11.19.3; 13.09.1)}

 \title{New Infrared Seyfert Galaxies}

 \author{Q.S. Gu\inst{1,3} \and J.H. Huang\inst{1,3} 
    \and H.J. Su\inst{2,3} \and Z.H. Shang\inst{2,3} } 

 \offprints{Q.S. Gu, e-mail:qsgu@nju.edu.cn}

 \institute{Department of Astronomy, Nanjing University, Nanjing 210093, China
 \and Purple Mountain Observatory, Nanjing 210008, China
 \and United Lab for Optical Astronomy, The Chinese Academy of Sciences, China}

 \date{Received ; accepted }

 \maketitle

\begin{abstract}
  We present optical spectra for new infrared Seyfert galaxies obtained with
  the 2.16m telescope at Beijing Astronomical Observatory (BAO). After 
  wavelength and flux calibration, they are classified by the degree of 
  nuclear activity: nine Seyfert 2 and three Seyfert 3 galaxies. 
  In addition, by using the data from de Grijp et al.(1992), we find that 
  (1) there exists a tight correlation between luminosities of 
  far-infrared (L$_{\rm FIR}$) and \Ha \ (L$_{\rm H\alpha}$) for both 
  Seyfert and HII-like (starburst) galaxies; (2) the median value of 
  \Ha \ luminosities of Seyfert 1s is one magnitude larger than that of 
  Seyfert 2s and starburst galaxies; (3) the cumulative distributions of 
  FIR luminosities and infrared spectral index $\alpha(100,60)$ for 
  Seyfert 1s and 2s are similar to that of starburst galaxies.  
  We conclude that most of the far-infrared emission from Seyfert 2 galaxies 
  is due to the violent nuclear/circumnuclear starburst, rather than the 
  nonthermal activity in the nucleus, this may also be the case for many 
  Seyfert 1s as well.


\end{abstract}

\section{Introduction}

 Seyfert galaxies have intense nuclear activities and they are strong sources
 of near and mid-infrared radiation (Rieke 1978). Since the IRAS survey 
 provided the infrared data for over 20,000 galaxies, it is a well 
 established fact that there is a strong 25$\mu$m component in Seyfert 
 galaxies (Dultzin-Hacyan et al. 1988); and de Grijp et al. (1985) presented 
 a new method for detecting hitherto unknown Seyfert galaxies by their flat 
 infrared spectra, where the infrared spectral indices between 25  $\mu$m 
 and 60 $\mu$m, $\alpha$(60,25) (defined by $S_{\nu} \propto \nu ^{\alpha}$, 
 where $S_{\nu}$ is the flux density at frequency $\nu$ ), are in the range 
 [-1.25, -0.5]. On the other hand, Keel et al. (1988) compared the 
 distribution of the location of stars, normal galaxies and AGN in the palne 
 of $\alpha$(60,25) vs $\alpha$(100,60) and found that AGN are located 
 in the area of $\alpha$(60,25) in the range [-1.5,0.5] or $\alpha$(100,60) 
 in the range [-0.8,0.5].

 Following de Grijp et al. (1985; 1992), we have selected a large sample 
 of galaxies to search for new Seyfert galaxies from
 the IRAS Extra-galactic Catalog (EGCAT, 1994) based on their IR properties.
 Since 1994, we have carried out the survey by using
 the 2.16m telescope at Beijing Astronomical Observatory (BAO).  
 Gu et al. (1995) presented the spectral results of ten new Seyfert 
 galaxies: one Seyfert 1, three Seyfert 2s and 6 LINERs (also called Seyfert 3 
 by Veron and Veron,1993).

 In this paper, we will present the spectral results of new Seyfert galaxies 
 which we detected in the recent observations. The organization of the paper 
 is as follows. In Sec.2, we describe our observations and data reduction 
 procedure, and we present the results of the twelve new Seyfert galaxies 
 in Sec.3. And in Sec.4, we present three diagnostic diagrams and discuss the 
 far-infrared (FIR) emission in Seyfert galaxies. Finally, the major results 
 of this paper are summarized in Sec.5.

\section{Observation and Data Reduction}

 The present mini-sample is selected with $\alpha$(60,25) in the range of 
 [-1.5,+0.5] and each galaxy has a strong 25 $\mu$m emission: that is the 
 flux ratio at 25 $\mu$m and 100$\mu$m is larger than 0.2. In order to be 
 sure that the objects are not foreground sources (stars, planetary 
 nebulae, etc), we have inspected each one using CDROM of the \DSS .

\begin{table*}
\caption[]{Observing log and some basic data}
\begin{flushleft}
\begin{tabular}{clllllllll}
\noalign{\smallskip}
\hline \noalign{\smallskip}
Object &  &Position & S12 & S25 & S60 & S100 &  Date \\
name    & $\alpha$(1950) & $\delta$(1950) &  Jy  & Jy & Jy  & Jy   & & \\
\hline \noalign{\smallskip}
 F01518+2705 & 015153.6 & +270505 & 0.189 & 0.690 & 1.224 & 1.755&Dec.17, 95\\
 F02095$-$0526 & 020933.9 & $-$052639 & 0.107&0.173&0.537 & 0.642&Dec.20, 95\\
 F02394+0457 & 023925.3 & +045724 & 0.043 & 0.159 & 0.555 & 1.734&Dec.18, 95\\
 F03077+1709 & 030746.9 & +170949 & 0.082 & 0.422 & 0.880 & 1.992&Dec.20, 95\\
 F04210+0401 & 042102.5 & +040105 & 0.084 & 0.453 & 0.919 & 1.491&Oct. 2, 95\\
 F04507+0358 & 045046.6 & +035847 & 0.362 & 0.599 & 0.612 & 0.041&Nov.28, 94\\
 F04580+0018 & 045800.3 & +001854 & 0.117 & 0.216 & 0.598 & 0.791&Dec.16, 95\\
 F08216+3009 & 082137.7 & +300907 & 0.215 & 0.375 & 0.441 & 0.354&Dec.17, 95\\
 F08449+3526 & 084457.6 & +352648 & 0.038 & 0.346 & 0.546 & 1.240&Jan.17, 96\\
 F09427+4252 & 094244.5 & +425222 & 0.177 & 0.290 & 0.663 & 0.946&Dec.20, 95\\
 F10285+5834 & 102834.3 & +583456 & 0.074 & 0.223 & 0.624 & 1.193&Dec.20, 95\\
 F10419+3430 & 104158.2 & +343045 & 0.050 & 0.226 & 0.574 & 0.647&Dec.16, 95\\
\noalign{\smallskip}
\hline
\end{tabular}
\end{flushleft}
\end{table*}

 The observing log and some basic data ( IRAS flux densities are taken from 
 EGCAT ) are presented in Table 1. All spectra were taken with the Carl 
 Zeiss Jena universal Cassegrain spectrograph at the Beijing Astronomical
 Observatory(BAO) 2.16m telescope, a 300 lines mm$^{-1}$ grating
 (dispersion is 195 \AA /mm, equivalently ) was used and the width of
 long slit was about 2.5".
 The spectral range was from 3800\AA \ to 7600\AA \ with the dispersion
 of 4.66 \AA/pixel and the resolution (FWHM) of 10.96 \AA. 
 Standard stars were selected  from KPNO standards, and all spectra 
 were reduced using standard \IRAF procedures.

\begin{figure}
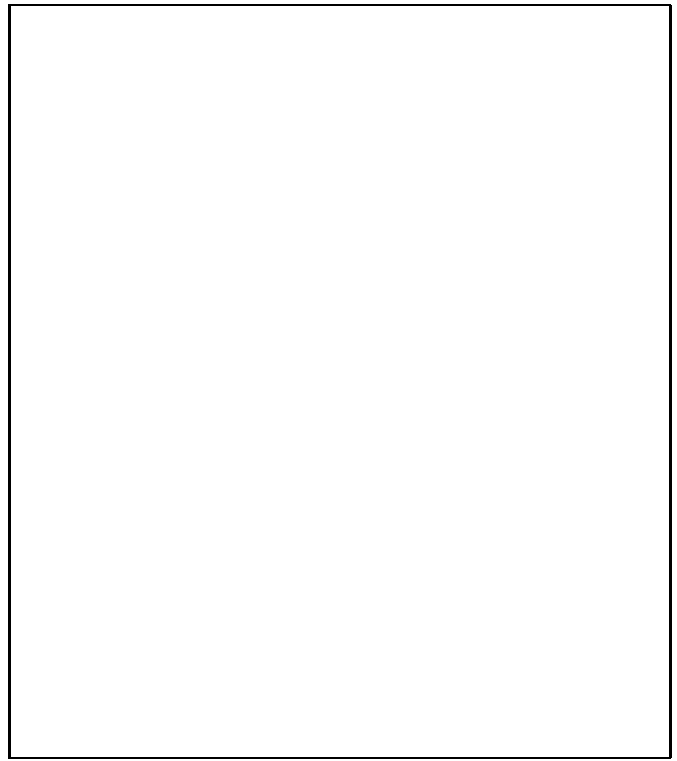

\picplace{10.cm}
\caption[]{The spectra of new Seyfert galaxies.  
           (a) F09427+4252, (b) F01518+2705.}
\end{figure}

\section{Results}

 The observed objects are classified according to their ionization states
 estimated from the flux ratios between emission lines (Baldwin et al. 1981;
 Veilleux \& Osterbrock 1987), the emission line ratios used are
 [OIII]$\lambda$5007/\Hb,   [NII]$\lambda$6583/\Ha ,
 [SII]$\lambda\lambda$6716,6731/\Ha \ and  [OI]$\lambda$6300/\Ha.

 Seyfert 1 galaxies have very broad HI, HeI and HeII emission lines
 with full widths at half maximum (FWHM) of the order of 1 to 5 $\times$
 10$^{3}$ km sec$^{-1}$, and the  forbidden lines such as 
 [OIII]$\lambda\lambda$4959,5007 and [NII]$\lambda\lambda$6548,6583 
 and [SII]$\lambda\lambda$6716,6731 have the FWHMs of order 500  
 km sec$^{-1}$ regardless of the narrow line ratios (Osterbrock 1989).
 The  objects with [OIII]$\lambda$5007/\Hb $>$3 and
 [NII]$\lambda$6583/\Ha $>$ 0.5 are classified as Seyfert 2, and those
 with [OIII]$\lambda$5007/\Hb $<$ 3 and [NII]$\lambda$6583/\Ha $>$0.5
 are classified as Seyfert 3.
 Emission-line objects whose line ratios lie outside the ranges 
 appropriate for Seyfert 2 and Seyfert 3 nuclei are
 called HII-like galaxies (de Grijp et al. 1992). The objects without
 [OIII]$\lambda$5007/\Hb, were classified as AGN-like
 if [NII]$\lambda$6583/\Ha $>$0.5 and [OI]$\lambda$6300/\Ha $>$0.06
 or as HII-like if [NII]$\lambda$6583/\Ha $<$0.5 and
 [OI]$\lambda$6300/\Ha $<$0.06 ( Armus, Heckman \& Miley, 1989 ).

 According to this classification scheme, we found nine Seyfert 2s and 
 three Seyfert 3s. We show the spectra of a typical Seyfert 2 and 
 Seyfert 3 in Figs. 1a and 1b.

\begin{table*}
\caption[]{Optical and infrared properties of the new Seyfert galaxies}
\begin{flushleft}
\begin{tabular}{ccrccccc}
\noalign{\smallskip}
\hline \noalign{\smallskip}
 IRAS Name & z & log L$_{\rm FIR}$ & [OIII]/\Hb  & [NII]/\Ha & [SII]/\Ha &
 [O I]/\Ha  & Spectral Type\\
(1) & (2) & (3) & (4) & (5) & (6) & (7) & (8) \\
\hline \noalign{\smallskip}
 F01518+2705 & 0.095 & 11.42  &2.4    &    0.947 &   0.479  & 0.162 &   Sy3 \\
 F02095$-$0526 & 0.041 & 10.31&      &    1.014 &   0.849  &       &   Sy2 \\ 
 F02394+0457 & 0.069 & 10.96  &4.553  &    0.869 &   0.573  &       &   Sy2 \\
 F03077+1709 & 0.065 & 11.03  &10.55  &    1.000 &   0.784  &       &   Sy2 \\
 F04210+0401 & 0.045 & 10.67  &13.393 &    0.302 &   0.637  & 0.168 &   Sy2 \\
 F04507+0358 & 0.028 &  9.88  &12.542 &    0.389 &   0.248  & 0.070 &   Sy2 \\
 F04580+0018 & 0.073 & 10.87  &       &    0.689 &   0.659  &       &   Sy3 \\
 F08216+3009 & 0.025 &  9.76  &12.90  &    0.338 &   0.199  & 0.107 &   Sy2 \\
 F08449+3526 & 0.057 & 10.71  &3.048  &    0.515 &   0.370  & 0.101 &   Sy2 \\
 F09427+4252 & 0.074 & 10.95  &5.068  &    0.516 &   0.292  & 0.047 &   Sy2 \\
 F10285+5834 & 0.091 & 11.15  &       &    0.811 &   0.815  &       &   Sy3 \\
 F10419+3430 & 0.070 & 10.80  &       &    0.656 &   0.361  &       &   Sy2 \\
\noalign{\smallskip}
\hline
\end{tabular}
\end{flushleft}
\end{table*}

\begin{figure*}
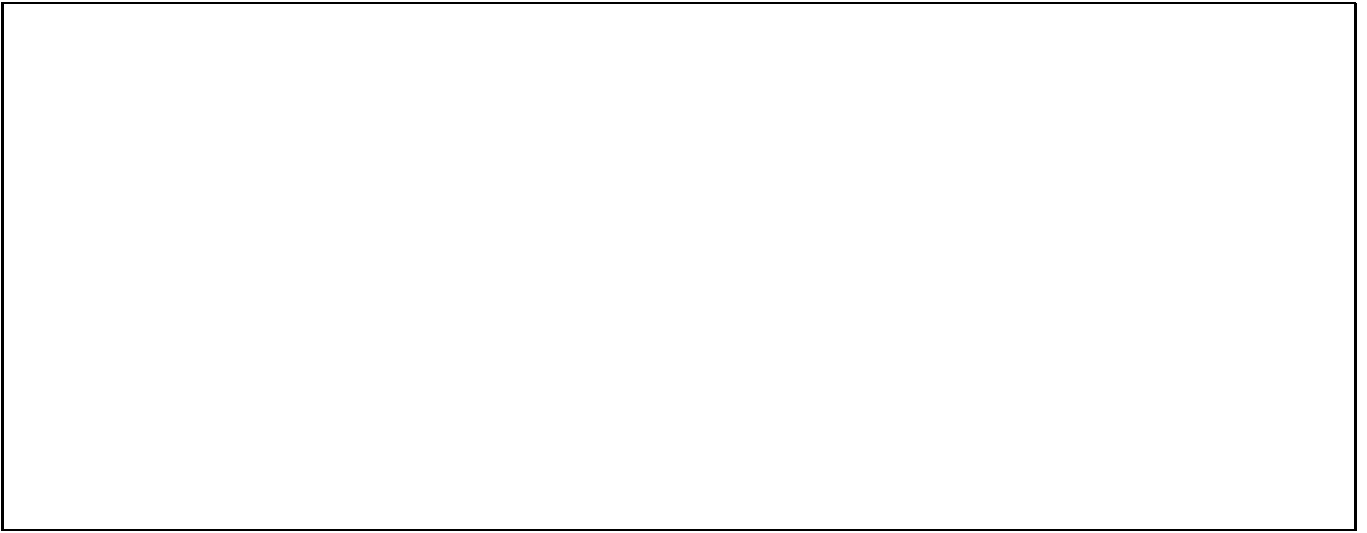

\picplace{7.cm}
\caption[]{Diagnostic diagrams .
 (a) log([NII]$\lambda$6583/\Ha) vs. log([OIII]$\lambda$5007/\Hb);
 (b) log([SII]$\lambda$6724/\Ha) vs. log([OIII]$\lambda$5007/\Hb);
 and (c) log([OI]$\lambda$6300/\Ha) vs. log([OIII]$\lambda$5007/\Hb).
 The short-dashed line divided NL AGNs from starburst galaxies is taken 
 from Osterbrock(1989). }
\end{figure*}

 The excitation states, along with the measured redshifts and far-infrared
 luminosities, of these new Seyferts are given in Table 2.
 The columns of Table 2 are as follows: column(1), the source name;
 column(2), the redshifts derived from \Hb, [OIII]$\lambda$5007,
 and \Ha; column(3), logarithm of the FIR luminosities in 
 \($L$_{\odot}\) from 40 to 120$\mu m$, given by Lonsdale et al. (1985)

\be
L_{\rm FIR}~=~3.75~10^{5}~D^{2}~(2.58~S_{60}+S_{100}) ,
\ee

\noindent
 where \(S_{60}\) and\(S_{100}\) are the flux densities at 60$\mu m$
 and 100$\mu m$ in Jy, D is the distance in Mpc
  and $H_{0} = 75 \ km \ sec^{-1} \ Mpc^{-1} $;
 column(4), the flux ratios of [OIII]$\lambda$5007 to \Hb;
 column(5), the flux ratios of [NII]$\lambda$6583 to \Ha;
 column(6), the flux ratios of [SII]$\lambda\lambda$6716,6731 to \Ha ;
 column(7), the flux ratios of [OI]$\lambda$6300 to \Ha;
 and column(8), the spectral type.

\section{Discussion}

\subsection{Diagnostic Diagrams}
 
 The diagnostic diagrams of the new Seyfert galaxies for which the 
 ratios [OIII]$\lambda$5007/\Hb, [OI]$\lambda$6300/\Ha, 
 [NII]$\lambda$6583/\Ha, and [SII]$\lambda$6724/\Ha \ could be measured ,
 are plotted in Figs.2a, b and c. All these new Seyferts occupy the 
 AGN portion of the plots where the photoionization by a power-law 
 continuum is the dominant ionization mechanism.

\begin{figure*}
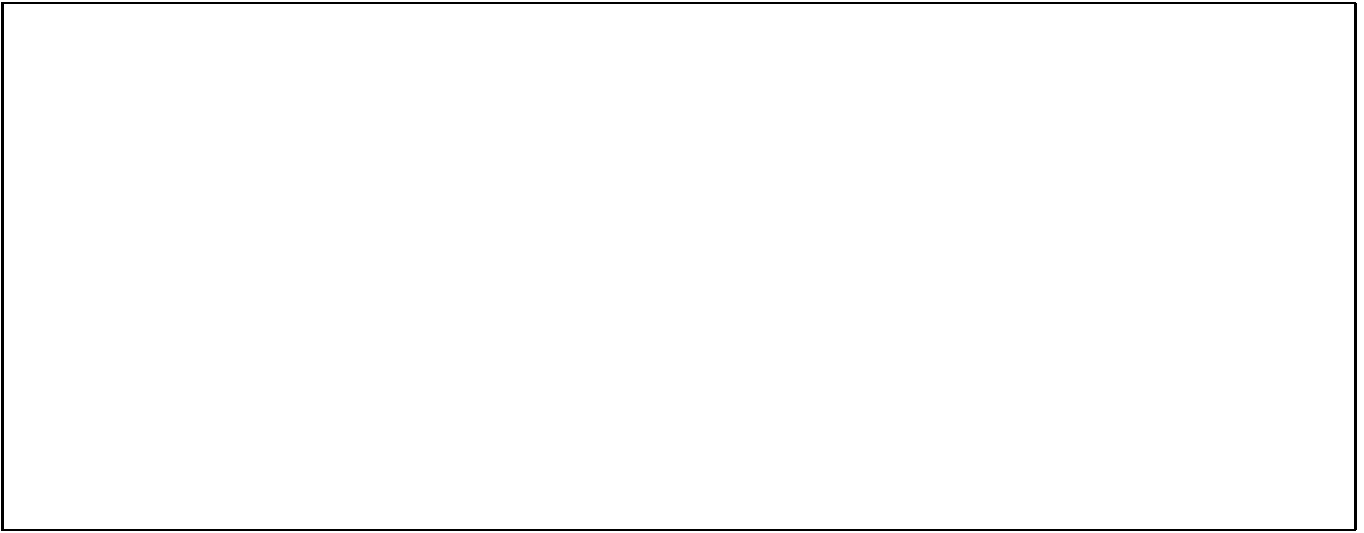

\picplace{7.cm}
\caption[]{Correlation between luminosities of far-infrared 
           (L$_{\rm FIR}$) and \Ha \ (L$_{\rm H\alpha}$). 
           (a) Seyfert 1 galaxies; (b) Seyfert 2 galaxies; 
           and (c) starburst galaxies. In (b), Seyfert 2 galaxies 
           taken from de Grijp et al.(1992) are plotted by open 
           circles and our new Seyfert 2s by filled circles.}
\end{figure*}

\subsection{FIR emission in Seyfert galaxies}

 FIR emission in Seyfert galaxies can be accounted for by (1) emission by
 warm dust which is heated in regions of star formation; (2) emission
 associated with an active galactic nucleus(AGN), either nonthermal flux
 coming directly from AGN or dust reradiation of nonthermal UV-optical
 continuum emission from the accretion disk. Most authors assume that FIR 
 emission in Seyferts comes from dust heated by power law continuum or is
 synchrotron radiation, e.g. Rowan-Robinson (1987).  
 But Rodriguez-Espinosa et al. (1987) analysed a sample
 of optically selected Seyfert galaxies and suggested that star formation 
 produced the bulk of the FIR emission in Seyfert galaxies.  
 In 1988, Dultzin-Hacyan et al. found a strong correlation between 
 the luminosity at 25 $\mu$m and the nuclear \Hb \ luminosity
 for Seyfert 2s but not for Seyfert 1s, so they suggested that only 
 Seyfert 2s have FIR emission from dust heated by hot stars, and they 
 also found the ratio of 25  $\mu$m to 100 $\mu$m for Seyfert 2s was 
 statistically equal to that for starburst galaxies, and suggested again  that
 hot stars are indeed capable of producing hot dust to emit observed  25 $\mu$m
 "excess" in Seyfert 2s (e.g. Vaceli et al., 1993, Mouri and Taniguchi, 1992).

 It is generally accepted that FIR emission in starburst galaxies 
 arises from the dust heated by newly formed OB stars and that \Ha \ 
 emission traces the stars producing significant ionizing radiation
 (OB stars) (Keel, 1991).

 In order to study FIR emission in infrared selected Seyfert galaxies, we 
 picked out all Seyfert and starburst galaxies from de Grijp et al. (1992), 
 together with our new Seyfert galaxies and those in Gu et al.(1995) to build 
 a large, complete infrared-selected Seyfert sample  and
 a starburst galaxies sample for comparison. The number of Seyfert 1s,
 Seyfert 2s, Seyfert 3s and starburst galaxies is 63(1), 141(12), 17(9) and 
 114(0), respectively, the digits in parenthesis are the number of our new
 Seyfert galaxies. We will not discuss the statistical properties of Seyfert 3 
 sample due to its small sample-size.

\begin{figure*}
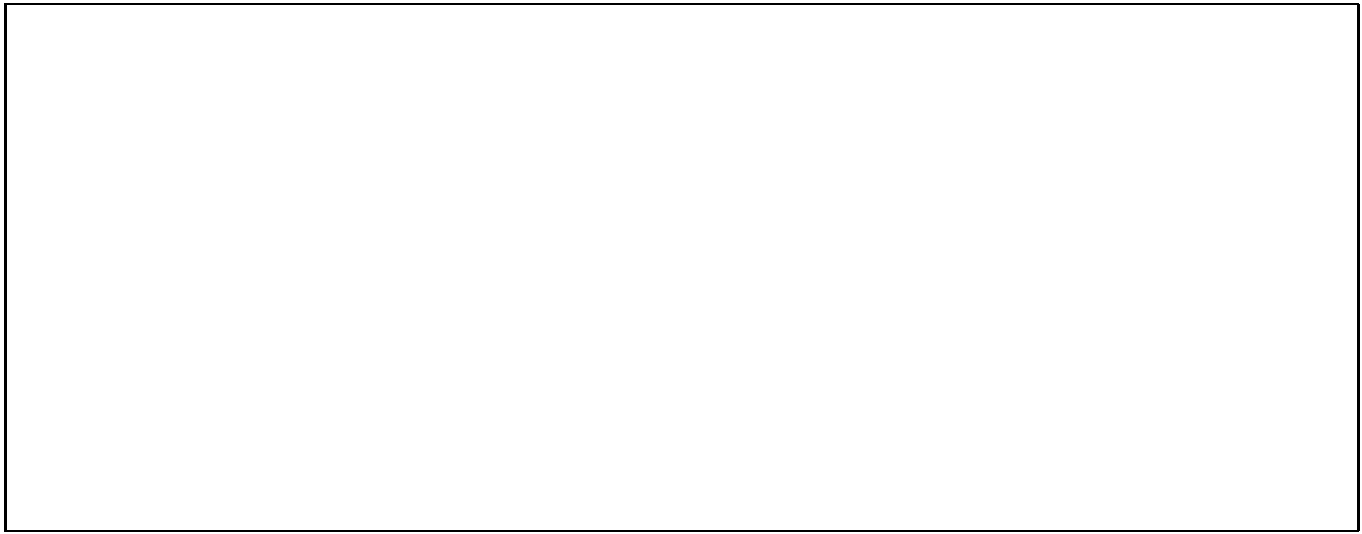

\picplace{7.cm}
\caption[]{Cumulative distributions of Seyfert 1, Seyfert 2, and starburst 
           galaxies.  (a) $L_{\rm H\alpha}$; (b) $L_{\rm FIR}$; and (c) IR 
           spectral index $\alpha(100,60)$.  Seyfert 1 galaxies are shown 
           by cross, Seyfert 2 by open circle and starburst galaxies by a 
           short bar. }
\end{figure*}

 Figs.3a, b  and c show the correlation between $L_{\rm FIR}$ and \Ha \ 
 luminosities ($L_{\rm H\alpha}$) for Seyfert 1s, Seyfert 2s and 
 starburst galaxies. The \Ha \ luminosity in logarithm is 
 computed by Devereux \& Young (1990),

\be
L_{\rm H\alpha} =  2.96 \times 10^{16} F_{\rm H\alpha} D^{2}  \ \  [L_{\odot}], 
\ee

\noindent
 where $F_{\rm H\alpha}$ is \Ha \ flux in units of ergs cm$^{-2}$ s$^{-1}$ 
 and D is the distance in Mpc.

 We also plot the regression line in each figure, which minimizes the sum 
 of the square of the perpendicular distances between data points and the 
 line (Isobe et al. 1990), the equations of three regression lines are as 
 follows:

\be
Syf \ 1s: log L_{\rm H\alpha} = (1.242 \pm 0.009) log L_{\rm FIR} - (4.506 \pm 1.322)
\ee

\be
Syf \ 2s: log L_{\rm H\alpha} = (1.191 \pm 0.006) log L_{\rm FIR} - (5.112 \pm 0.883)
\ee

\be
SBs : log L_{\rm H\alpha} = (1.050 \pm 0.005) log L_{\rm FIR} - (3.555 \pm 0.669)
\ee
 
 The correlation coefficients for Seyfert 1s, 2s and starburst galaxies 
 are 81.7 \%, 66.7 \% and 66.0 \%, respectively. It is shown that there
 is a tight correlation between $L_{\rm FIR}$ and $L_{\rm H\alpha}$ in 
 starburst galaxies which has been predicted theoretically as both FIR 
 and \Ha \ emission arise from star formation region. 
 But such tight correlation also exists in Seyfert galaxies, and even 
 better than that in starburst galaxies. We notice that the non-linearity 
 is becoming obvious from 1.050 in starburst to 1.191 in Seyfert 2s and 1.242 
 in Seyfert 1 galaxies, such change of the slope may reflect the increasing 
 contribution to FIR emission from the nonthermal activity in AGN. 

 We also show the cumulative distributions of $L_{\rm H\alpha}$ , $L_{\rm FIR}$
 and IR spectral index $\alpha(100,60)$ in Figs.4 a, b and c for Seyfert 
 1s(cross), 2s(open circle) and starburst galaxies(-). The median values of 
 these three parameters for Seyfert 1, 2 and starburst galaxies are  given 
 in Table 3.

\begin{table}
\caption[]{Median values of $L_{\rm H\alpha}$, $L_{\rm FIR}$ and 
         $\alpha(100,60)$ for Seyfert 1s , 2s and starburst galaxies}
\begin{tabular}{lccc}
\hline
Sample & $log L_{\rm H\alpha}$ & $log L_{\rm FIR}$ & $\alpha(100,60)$ \\
       & $L_{\odot}$       & $L_{\odot}$   &                  \\
\hline
 Seyfert 1 & 8.467         & 10.368         &  -0.826  \\
 Seyfert 2 & 7.423         & 10.532         &  -0.623 \\
 starburst  & 7.335         & 10.392         &  -0.780 \\
\hline
\end{tabular}
\end{table}

 From Fig4b. and 4c, we find that the distributions of $L_{\rm FIR}$ and 
 $\alpha(100,60)$ are both similar for these three samples which indicate 
 that the circumnuclear starburst is energetic enough to account for FIR 
 emission in Seyfert galaxies. It is very interesting to note that 
 $L_{\rm H\alpha}$ of Seyfert 1 galaxies is 
 about one magnitude larger than that of Seyfert 2 and starburst galaxies.
 On the other hand, the distribution for Seyfert 2 is almost the same as 
 the one for starburst galaxies.  In the view of the unified scheme of 
 AGN (see Antonucci 1993), Seyfert 1 and Seyfert 2 
 galaxies share the same nuclei and the observed differences are due to
 obscuration and viewing angle effect and not to intrinsic, physical
 differences.  According to this scheme, the \Ha \ emission in Seyfert 2s 
 from the circumnuclear starburst suffered severe attenuation by the dusty
 torus around the accretion disk while not in Seyfert 1s. So the observed
 systematical difference  in \Ha \ emission  between Seyfert 1s and 
 Seyfert 2s, shown in Fig. 4a, may give another evidence supporting 
 the unified scheme. The optical depth ($\tau_{d}$) of this dusty torus
 could be inferred in the following. Assuming the observed 
 \Ha \ intensity from Seyfert 2 to be, 
 $I_{\rm 2,obs} = I_{0} e^{ - \tau_{d}}$, then $\tau_{d}$ is approximate to
 $\ln (10^{8.467-7.423}) = 2.4$.

 Now, we could estimate the thermal contribution to FIR emission 
 from the intense circumnuclear starburst in Seyfert 2 galaxies 
 by a very simplified approach. 
 Assuming that the FIR emission in starburst galaxies completely comes from 
 starburst, the FIR-emission percentage  contributed by starburst in 
 Seyfert 2s is about $ 10^{(\frac{7.423-7.335}{1.050}+10.392-10.532)} 
 = 88 \% $.

 Dultzin-Hacyan et al. (1990) have convincingly indicated that the ratio of
 $I_{25}/I_{100}$ is the best IRAS tracer of recent star formation.
 We found that the mean value of $I_{25}/I_{100}$ for Seyfert 2s,
 0.3589, is statistically equal to that of Seyfert 1s, 0.3637.
 So hot stars can also heat dust up to the observed temperature 
 for 25 $\mu$m "excess" in Seyfert 1s. It is interesting to notice that 
 this mean value of $I_{25}/I_{100}$ for IR selected Seyferts is
 statistically the same as that of optically selected Seyfert 1s
 (Dultzin-Hacyan et al. 1988).
 Our results confirm previous claims that FIR comes from dust heated by hot
 stars in Seyfert 2s and we also find that it is the same for Seyfert 1s in
 our IR selected sample, and possibly for optically selected Seyfert 1s, too.

\section{Conclusion}
 We have presented the results of the slit-spectroscopic observations 
 of 12 new infrared-selected Seyfert galaxies which are classified according
 to their principle excitation mechanisms as following: nine of them to be 
 Seyfert 2 galaxies and three to be Seyfert 3 galaxies.
 
 By comparing the properties of IR selected Seyfert and starburst galaxies, we 
 obtained the following results:

 1. There is a tight correlation between FIR and \Ha \ luminosities for 
    Seyfert and starburst galaxies , which indicates that FIR emission in 
    both Seyfert and starburst galaxies arises from starburst activities.

 2. The median value of \Ha \ luminosities of Seyfert 1s is one magnitude 
    larger than that of Seyfert 2s and starburst galaxies, which is 
    consistent with the AGN unified scheme.

 3. The cumulative distribution of $L_{\rm FIR}$ for Seyfert galaxies is 
    similar to that of starburst galaxies. A circumnuclear starburst in 
    Seyfert galaxies is energetic enough to account for FIR emission.

 4. The cumulative distribution of IR spectral index $\alpha(100,60)$ for 
    Seyfert galaxies is also similar to that in starburst galaxies, showing 
    that FIR emission in Seyfert galaxies comes from dust reprocessing 
    emission, corresponding to the same temperature as that in starburst 
    galaxies.

 Our work gives strong support to earlier claims that the FIR luminosity of 
 Seyfert 2 galaxies is of stellar origin (e.g. Rodriguez-Espinosa et al..,1987; 
 Mouri \& Taniguchi 1992;  Dultzin-Hacyan \& Benitez 1994). Moreover, here 
 we show that this may also be the case for many Seyfert 1 galaxies as well. 
 At least, it is the case for all FIR selected Seyfert 1s in our sample. It 
 is important to notice, however, that our sample is biased towards galaxies 
 with recent bursts of star formation, because it is FIR selected and because
 all galaxies have relative flat infrared spectra 
 (see Dultzin-Hacyan, Masegosa \& Moles 1990).

\begin{acknowledgements}
 We are very grateful to the referee, Dr. D. Dultzin-Hacyan, for her
 careful reading the manuscript and many constructive comments which 
 improved the paper a lot. We also wish to thank Dr. J.M. Wang for 
 helpful discussion and to the staff of the observatory at BAO for 
 their supports during the observing run. This work has been supported 
 by grants from National Science and Technology Commission and National 
 Natural Science Foundation of China.
\end{acknowledgements}

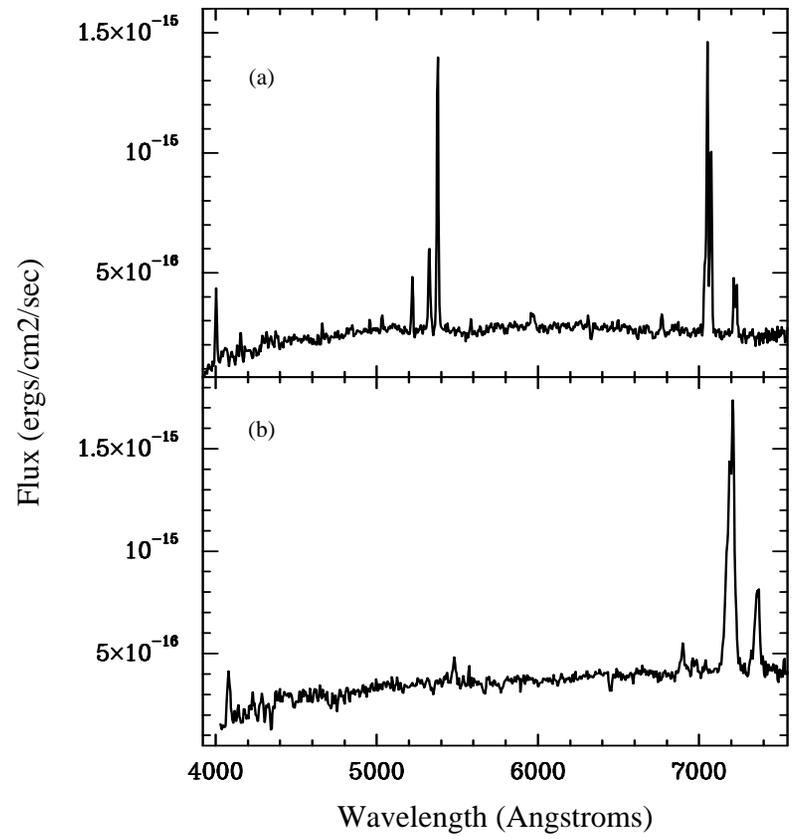
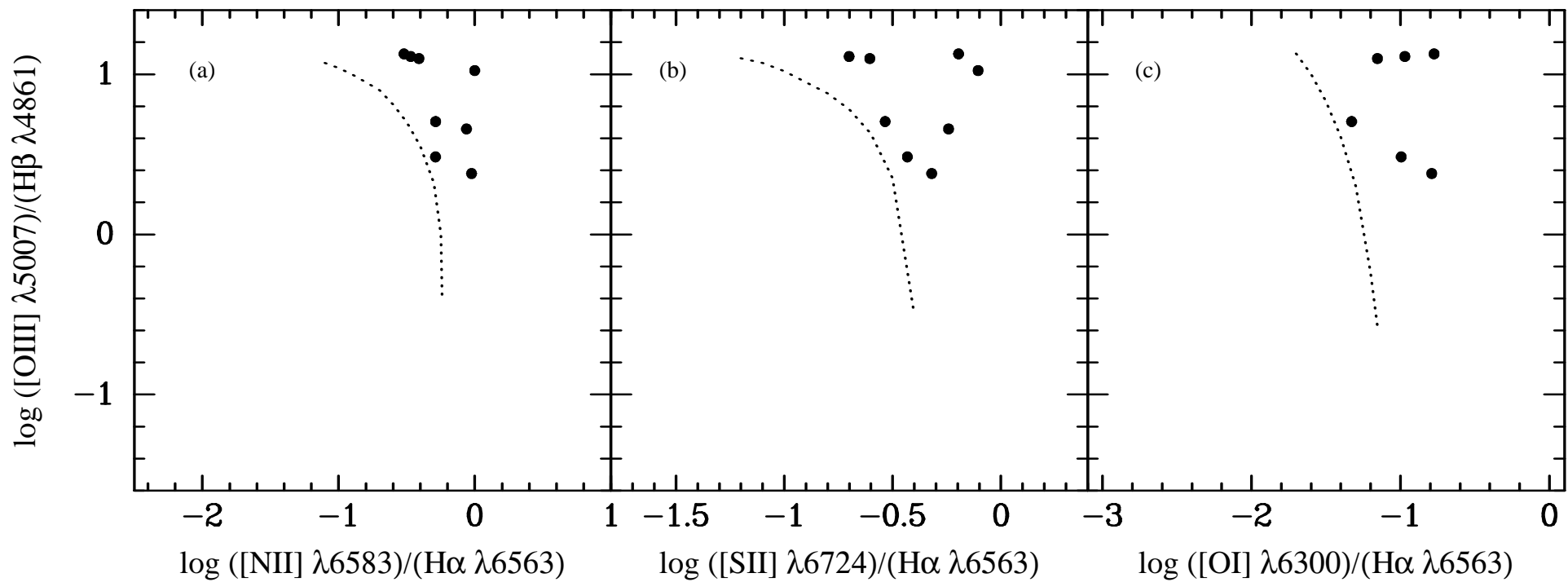
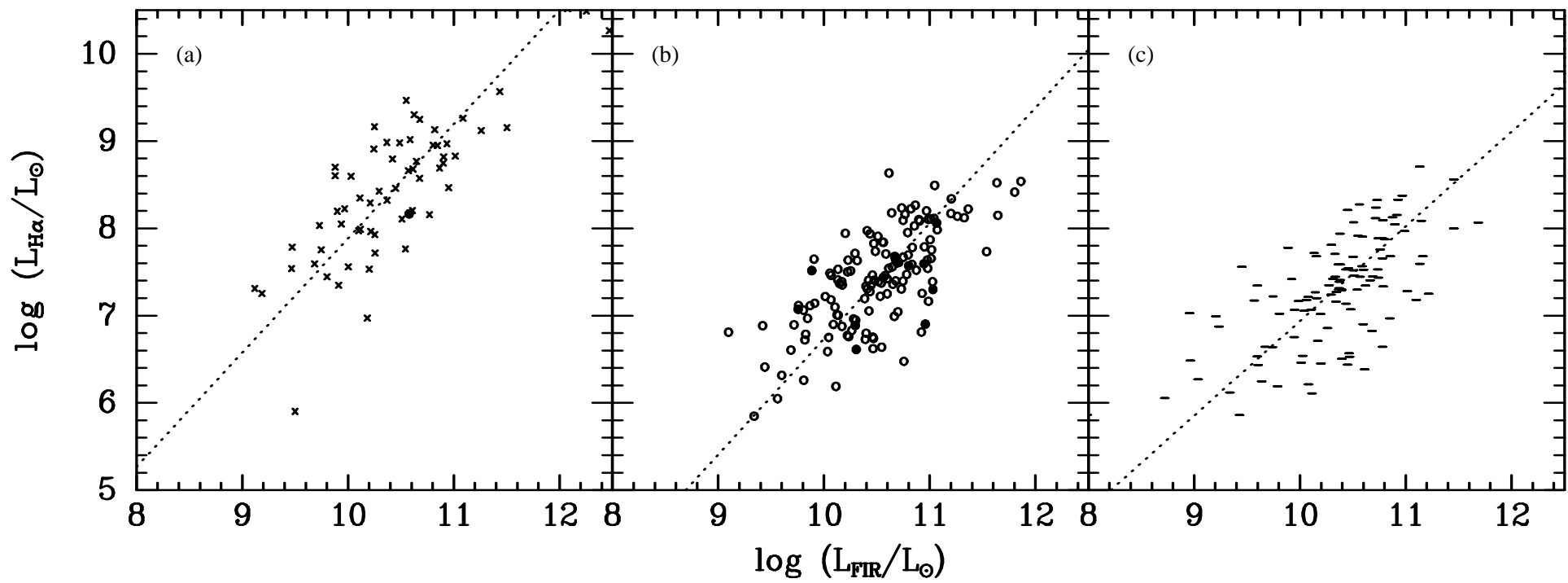
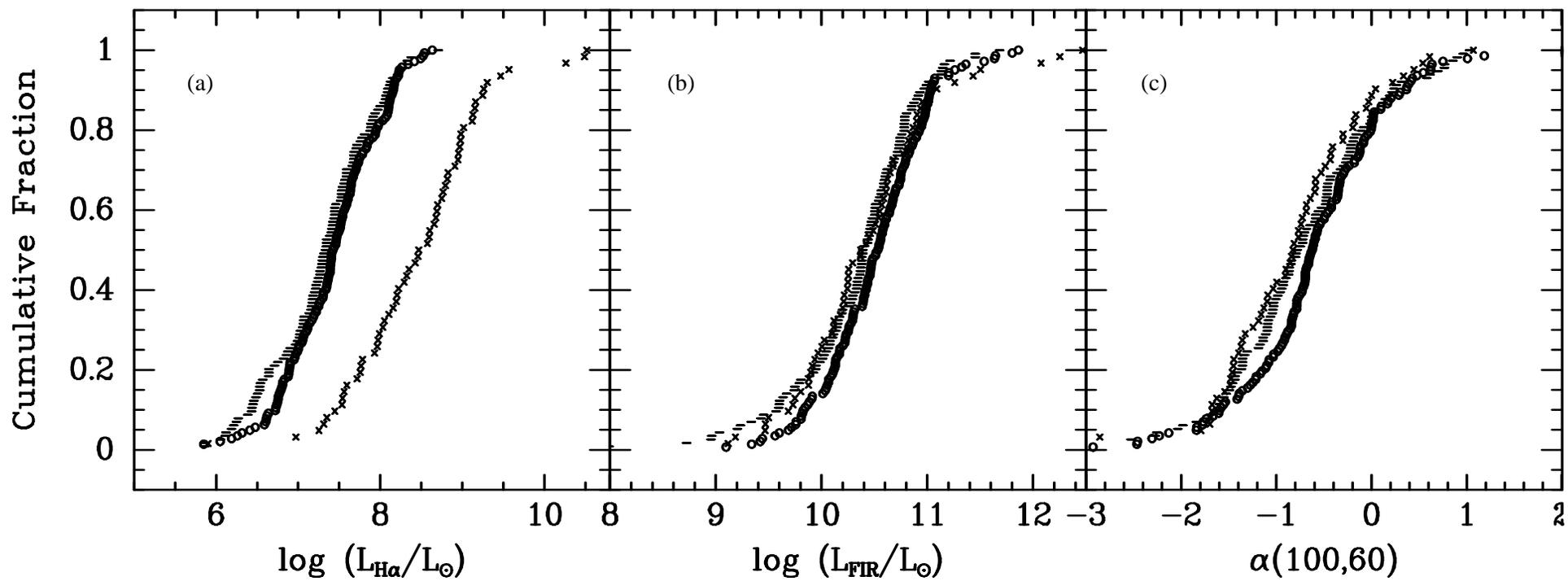
\end{document}